\documentclass[prb,aps,amsmath,amssymb,superscriptaddress,twocolumn]{revtex4}
\usepackage{graphicx}
\usepackage{dcolumn}
\usepackage{enumitem}  
\usepackage{amsmath}

\begin{document}
\title{Magnetic properties of the frustrated decorated Ising chain.}

\author{D.V. Laptiev}
\affiliation{B.I. Verkin Institute for Low Temperature Physics and Engineering of the National Academy of Sciences of Ukraine, Nauky Ave., 47, Kharkiv, 61103, Ukraine}
\email{laptev@ilt.kharkov.ua}

\author{O.O. Krivchikov}
\affiliation{B.I. Verkin Institute for Low Temperature Physics and Engineering of the National Academy of Sciences of Ukraine, Nauky Ave., 47, Kharkiv, 61103, Ukraine}

\author{Yu.V. Savin}
\affiliation{B.I. Verkin Institute for Low Temperature Physics and Engineering of the National Academy of Sciences of Ukraine, Nauky Ave., 47, Kharkiv, 61103, Ukraine}
\affiliation{Simon Kuznets Kharkiv National University of Economics, Nauky Ave., 9-A, Kharkiv, 61166, Ukraine}

\author{V.V. Slavin}
\affiliation{B.I. Verkin Institute for Low Temperature Physics and Engineering of the National Academy of Sciences of Ukraine, Nauky Ave., 47, Kharkiv, 61103, Ukraine}
\affiliation{V.N. Karazin Kharkiv National University, 4, Svoboda sq. Kharkiv, 61022, Ukraine} 
\begin{abstract}
Using the Kramers-Wannier transfer matrix method we studied several decorated Ising chains. The exact expressions for thermodynamic characteristics, including the ground state characteristics, were  obtained. We considered a number of modeling chains with different signs and absolute values of exchange constants for the nearest- and the next-nearest neighbors. 
For these models we calculated the magnetization curves. The critical values of magnetic fields and corresponding magnetization plateau parameters were obtained. 
Analytic expressions for the ground state entropy were obtained for the chains with different interaction constants. The dependencies of the number of states with minimum energy (the degeneration of the ground state) as the function of the number of particles were found. It was shown that these dependencies are expressed in terms of well-known numerical sequences - Lucas numbers and Pell numbers, which, in the limit of a large number of particles, are proportional to the powers of the golden and silver sections. Therefore, the ground state entropy (per particle) of the systems under consideration can be described in terms of these sections and, therefore, is nonzero.   
\end{abstract}
\date{\today}
\maketitle

\section{Introduction}
\begin{figure}
\center{\includegraphics[width=8cm]{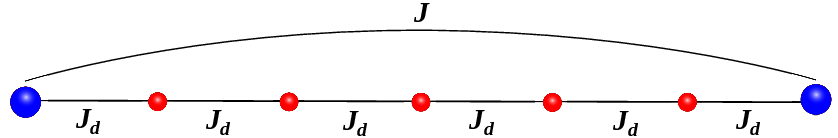}}
\caption{Decorated chain with exchange interaction of nearest neighbors $J_d$ and exchange interaction $J$ \cite{lit10}.}
\label{fig1}
\end{figure}

In recent years, the investigation of decorated spin lattices hase become an important area of in the physics of magnetic phenomena \cite{lit1,lit2,lit3,lit4}. 
This is due to the fact that lattice decoration generates a number of new effects that have not yet been fully studied. The term decorated lattice was introduced by I. Syozi \cite{lit5}, 
and later this concept began to develop rapidly \cite{lit6,lit7,lit8,lit9,lit10}. 
Decoration of lattice consists in introducing an extra (decorating) spin on every bond between the sites of the original lattice (nodal spins). 
Syozi also introduced a special transformation (decoration–iteration transformation) which allows us to derive the thermodynamic properties of the decorated lattice from those of the original undecorated one.  Figure \ref{fig1} shows an example of a one-dimensional decorated chain \cite{lit10}. Spins that participate in the interaction between nearest neighbors, as well as in interaction with neighboring spins located through one, two, or generally speaking d nodes, are called nodal spins (blue circles in Fig.~\ref{fig1}), the remaining spins are called non-nodal (decorated) spins 
(red circles in Fig.~\ref{fig1}). The index $d$ denotes the number of so-called chain decorations.

It is worth emphasizing that the vast majority of real structures are decorated. Moreover, some crystalline compounds can also be considered as decorated lattices. For example, fcc lattices are decorated along six faces of a cube (face-decorated cubic lattice), bcc lattices are decorated along the volume of a cube (body-decorated cubic lattice).
The investigation of the decorated spin lattices with frustrations are of particular interest due to wide spectrum of unusual effects inherited by frustrated systems. However, the complete theoretical explanation of number of experimental results, observed in these systems, are absent up to now. For example, the construction of phase diagrams for high decoration orders has not yet been done \cite{lit15}.

\section{The model and Hamiltonian}
\label{Sec:model}
\begin{figure}[ht]
\center{\includegraphics[width=8cm]{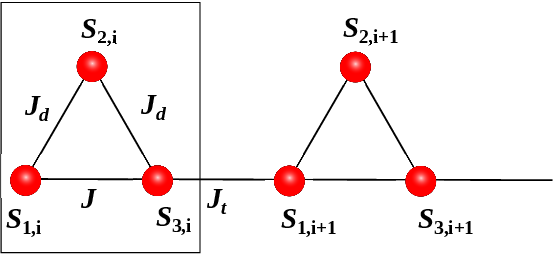}}
\caption{1D decorated Ising chain (the unit cell is inside a rectangular box). }
\label{fig:model}
\end{figure}

We consider one dimensional (1D) decorated Ising chain, presented in Fig.~\ref{fig:model} in the external magnetic field $h$. 
We assume that the number of spins in the lattice $N$ is multiple of $3$, so the system can be represented as a chain of an integer number of spins triangles. 
Note that all the quantities will be considered in the dimensionless units: Boltzmann constant $k_B=1$, Bohr magneton $\mu_B=1$, Lande factor $g=1$.

The Hamiltonian of the system consists of two terms:

\begin{equation}
H=H_0+H_h
\label{Ham1}
\end{equation}

The first term $H_0$ corresponds to the interaction between the spins:

\begin{equation}
H_0=-\sum_{i=1}^{N/3} J_d (S_{1,i}S_{2,i} +S_{2,i}S_{3,i}) +J S_{1,i}S_{3,i} +J_t S_{3,i}S_{1,i+1}\, .
\label{H_0}
\end{equation}

Here the first index in $S_{k,i} =\pm 1$ enumerates spins in unit cell ($k=1,2,3$) and the second index enumerates the unit cells ($i=1,2,\ldots, N/3$). 
The exchange interaction constants $J_d$, $J_t$, $J$ are shown in the Fig.~\ref{fig:model}. $J_d$ corresponds to the exchange interaction between the nearest decorating spins and 
nodal spins and between the nearest decorating spins, $J$ -- corresponds to the exchange interaction between the nearest nodal spins. 
$J_t$ corresponds to the exchange interaction between two nearest spins from two adjacent spin ``triangles''. 

The second term $H_h$ is the interaction energy of the spins with the external magnetic field $h$:

\begin{equation}
H_h=-h\sum_{i=1}^{N/3} (S_{1,i} + S_{2,i} + S_{3,i}).
\label{H_h1}
\end{equation}

\begin{figure}[ht]
\center{\includegraphics[width=8cm]{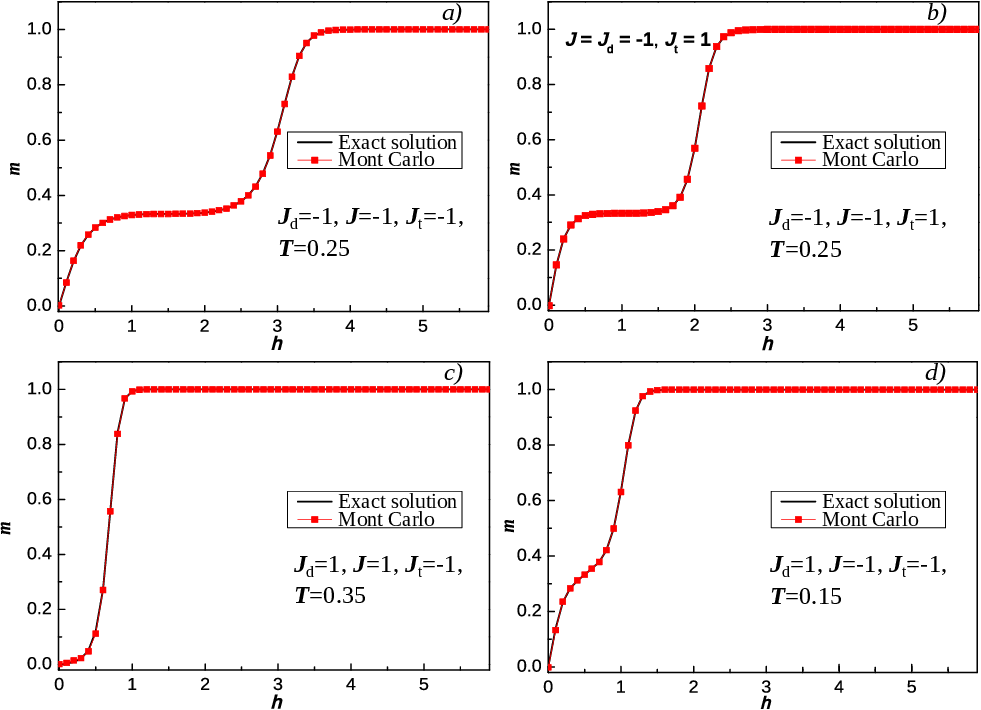}}
\caption{Low temperature field dependencies of the magnetization $m$ at different temperatures and different values of exchange constants.
a) $J_d=-1$, $J=-1$, $J_t=-1$, $T=0.25$;
b) $J_d=-1$, $J=-1$, $J_t=1$,  $T=0.25$;
c) $J_d=1$, $J=1$, $J_t=-1$,   $T=0.35$;
d) $J_d=1$, $J=-1$, $J_t=-1$,  $T=0.15$.
}
\label{fig:m_h_low_T}
\end{figure}

\section{Thermodynamics of the decorated Ising chain}

The partition function of the system under consideration is:

\begin{equation}
Z = \sum_{\{S =\pm 1\}} e^{-\beta H} \, .
\label{BigSum}
\end{equation}
\noindent Here $\{S =\pm 1\}$ means summation over all spin variables $\{S_{k,i}\}_{k,i=1}^{3,N/3}$; $\beta = 1/T$ ($T$ is the temperature). 
Imposing the periodic boundary conditions 

\begin{equation}
S_{1,N+1} = S_{1,1},
\label{Boundary_cond}
\end{equation}

\noindent one can represent (\ref{H_h1}) in symmetric form:

\begin{equation}
H_h= -h \sum_{i=1}^{N/3} \frac{S_{1,i} + S_{2,i}}{2} + \frac{S_{2,i} + S_{3,i}}{2} + \frac{S_{3,i} + S_{1, i+1}}{2} .
\label{H_h2}
\end{equation}

\noindent and rewrite (\ref{BigSum}) as the following:

\begin{equation}
Z = \sum_{\{S=\pm 1\} }  \prod_{i=1}^{N/3} V_i(S_{1,i}, S_{2,i}, S_{3,i}, S_{1, i+1})\, ,
\label{BigSum1}
\end{equation}

\noindent where

\begin{flalign}
& V_i(S_{1,i}, S_{2,i}, S_{3,i}, S_{1, i+1}) = &  \nonumber \\
& \exp(-\beta\{J_dS_{2,i}(S_{1,i} +  S_{3,i}) + J S_{1, i} S_{3,i} + J_t S_{3,i}S_{1, i+1}  + & \nonumber \\ 
& \frac{h}{2}[ (S_{1,i} + S_{2, i}) + (S_{2,i} + S_{3,i}) + (S_{3,i} + S_{1, i+1}) ] \} ) 
\label{V_i}
\end{flalign}

\begin{figure}
\center{\includegraphics[width=8cm]{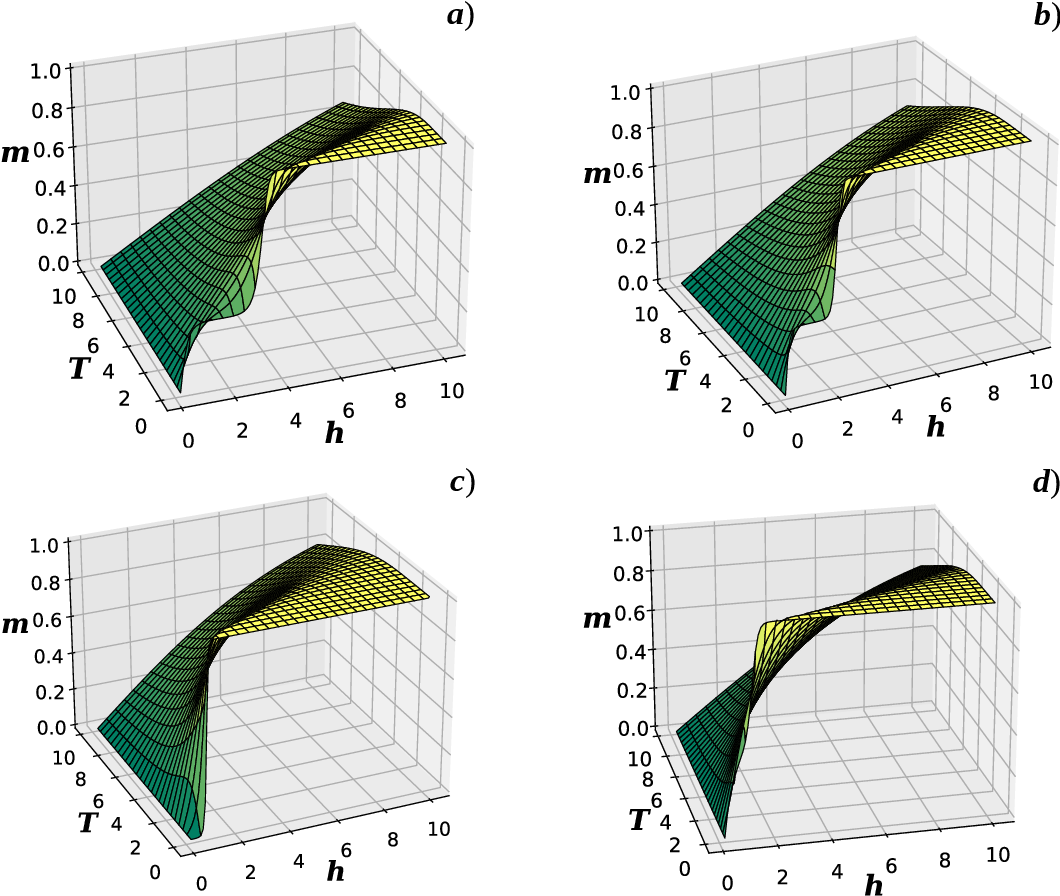}}
\caption{Magnetization $m$ as the function of magnetic field $h$ and temperature $T$ at different values of exchange constants.
a) $J_d=-1$, $J=-1$, $J_t=-1$;
b) $J_d=-1$, $J=-1$, $J_t=1$;
c) $J_d=1$, $J=1$, $J_t=-1$;
d) $J_d=1$, $J=-1$, $J_t=-1$.
}
\label{fig:m_h_and_T}
\end{figure}

Due to translational invariance of the system under consideration one can omit the second index for spins variables $S_{k,i}$ and, hence, the
expression (\ref{BigSum1}) acquires the form:

\begin{equation}
Z = \sum_{S_1, S_2, S_3, S_4=\pm 1} V^{\frac{N}{3}} (S_{1}, S_{2}, S_{3}, S_{4})
\label{BigSum2}
\end{equation}

\noindent where  $S_4=S_{1,i+1}$ and 


\begin{flalign}
& V(S_1, S_2, S_3, S_4) =   \nonumber \\
& \exp(-\beta\{J_dS_{2}(S_{1} +  S_{3}) + J S_{1} S_{3} + J_t S_{3}S_{1}  +  \nonumber \\ 
& \frac{h}{2} (S_{1} + S_{2} + S_{3} +  S_{4}) \} ) 
\label{V_2}
\end{flalign}

\noindent Performing summation over ``inner'' variables $S_2$ and $S_3$ one
can rewrite the expression (\ref{BigSum2}) as the trace of $2\times2$ matrix \cite{lit18}:

\begin{equation}
Z = \rm{Tr}\left[ R^{N/3} \right] = \lambda_+^{N/3}  + \lambda_-^{N/3}
\label{BigSum3}
\end{equation}

\noindent where

\begin{equation}
R = \left\{ R(\alpha, \beta)\right\}_{\alpha,\beta = \pm 1} = \sum_{S_2,S_3=\pm 1} V(\alpha, S_2, S_3, \beta)
\label{TransferMatrix}
\end{equation}

\noindent is the transfer matrix with the following matrix elements:
\begin{flalign}
R(+,+) = & 2\exp(x+2h') \cosh(z+h')+\nonumber\\
		 & 2\exp(-x) \cosh(h') \nonumber \\
R(+,-) = & 2\exp(y+h') \cosh(z+h')+ \nonumber \\
		 & 2\exp(-y-h') \cosh(h') \nonumber \\
R(-,+) = & 2\exp(y-h') \cosh(z-h')+ \nonumber \\
		 & 2\exp(-y+h') \cosh(h')\nonumber \\
R(-,-) = & 2\exp(x-2h') \cosh(z-h')+ \nonumber \\
		 & 2\exp(-x) \cosh(h'), 
\end{flalign}

\noindent (here $x=\beta(J+J_t)$, $y=\beta(J-J_t)$, $z=2\beta J_d$, and $h'= \beta h$). 

$$\lambda_\pm = \frac{1}{2} \left({\rm Tr}[R] \pm \sqrt{\left({\rm Tr}[R]\right)^2 - 4 {\rm Det}[R]} \right)$$ 
are the  eigenvalues of $R$. The explicit expressions for $\lambda_\pm$ are very cumbersome and for this reason we do not mention them in our work. 

In thermodynamic limit ($N\to\infty$) we have
$$Z = \lambda^{N/3}.$$
\noindent Here $ \lambda=\lambda_+$. The corresponding quantitatives (per particle): free energy $f$, entropy $s$ and magnetization $m$ are determined by the 
regular expressions (see (\ref{Free_Energy}), (\ref{Entropy}) and (\ref{Magnetization})).

\begin{equation}
f = -\frac{T}{N} \ln(Z)=-\frac{T}{3}\ln(\lambda) 
\label{Free_Energy}
\end{equation}
\begin{equation}
s = -\frac{\partial f}{\partial T}=\frac{1}{3}\left(\ln(\lambda) + \frac{T}{\lambda} \frac{\partial \lambda}{\partial T} \right)
\label{Entropy}
\end{equation}
\begin{equation}
m = -\frac{\partial f}{\partial h}= \frac{T}{3\lambda}\frac{\partial \lambda}{\partial h} 
\label{Magnetization}
\end{equation}

\section{Magnetic and thermodynamic properties of decorated Ising chain}
\label{sec:mag_therm}
Despite the simplicity, the system under consideration demonstrates rather interesting and rich in content thermodynamic behavior.
This behavior depends qualitatively on the values (including signs) of exchange parameters $J_d$, $J$, and $J_t$. 
Here we will consider several cases, corresponding to the most interesting cases, which are qualitatively different. 

\begin{enumerate}[label=\alph*)]
\item $\{J_d=-1$, $J=-1$, $J_t=-1\}$  --- ``antiferromagnetic'' interaction inside triangles, ``antiferromagnetic'' interaction between triangles.
\item $\{J_d=-1$, $J=-1$, $J_t=1\}$   --- ``antiferromagnetic'' interaction inside triangles, ``ferromagnetic'' interaction between triangles.
\item $\{J_d=1$, $J=1$, $J_t=-1\}$   --- ``ferromagnetic'' interaction inside triangles, ``antiferromagnetic'' interaction between triangles.
\item $\{J_d=1$, $J=-1$, $J_t=-1\}$   --- mixed ``ferromagnetic''/``antiferromagnetic'' interactions inside triangles, ``antiferromagnetic'' interaction between triangles.
\end{enumerate}

To check our results we performed an addition classic Monte Carlo simulation of magnetization $m$ as the function of external field $h$ for different combination of exchange constants
and different values of temperature $T$. The results are presented in Fig.~\ref{fig:m_h_low_T}. Black curves were obtained using (\ref{Magnetization}), red curves correspond to Mont Carlo simulations.
One can see good agreement between the results.
 
Due to ``antiferromagnetic'' interaction inside triangles (i.e. due to frustrations), $m=1/3$ plateau of intermediate magnetization is observed (see Fig.~\ref{fig:m_h_low_T}a  and Fig.~\ref{fig:m_h_low_T}b). 
Analogically, due to frustrations inside triangles $m=1/3$ plateau exists in the case d) also. 
This plateau corresponds to the spin configurations, where the only one spin in each triangle is oriented along magnetic field.
On the contrary, in the system with ``ferromagnetic'' interaction inside triangles the spins of triangles play the role of a ``effective 
spin'' and the magnetization curves are typical for 1D  Ising chain with ``antiferromagnetic'' interaction between triangles  (see Fig.~\ref{fig:m_h_low_T}c)). 

On Fig.~\ref{fig:m_h_and_T} the dependencies of the magnetization $m$ on magnetic field $h$ and temperature $T$ for the same combination of exchange constant are shown.
Using these dependencies one can obtain, in particular, the critical values of magnetic fields $h_c$  --- quantum ($T=0$) phase transition fields  
and the residual magnetization 
\begin{equation}
m_0=m(h\to+0, T=0)\ .
\label{m0}
\end{equation}
\noindent  As it seen $m_0=1/3$ in the cases cases {\it a, b, d} and  $m_0=0$ in the case {\it c}.
The critical values of magnetic field $h_c$ and the residual magnetization values $m_0$ are presented in Table~\ref{Table1}.
Note, that exact values of $h_c$ were obtained analytically from the expression (\ref{Magnetization}) in the limit $T\to 0$. 

\begin{table}[ht]
\begin{tabular}{|c|l|c|c|}
\hline
Model & Interaction constants & $h_c$ & $m_0$ \\
\hline
a) & $\{J_d=-1$, $J=-1$, $J_t=-1\}$ & $3$ & $1/3$ \\
\hline
b) & $\{J_d=-1$, $J=-1$, $J_t=1\}$ & $2$ & $1/3$ \\
\hline
c) & $\{J_d=1$, $J=1$, $J_t=-1\}$  & $2/3$& $0$ \\
\hline
d) & $\{J_d=1$, $J=-1$, $J_t=-1\}$ & $1$& $1/3$\\
\hline
\end{tabular}
\caption{Critical magnetic field for different values of exchange constants.}
\label{Table1}
\end{table}

These dependencies, in particular, allows us to establish the dependence of magnetization $m$ on $h$ at $h=h_c$ as the function of temperature $T$ in the limit $T\to 0$. 
In this limit we obtained the zero temperature magnetization at $h=h_c$: $m_c=m(h=h_c,T=0)$.
It allows us to extend the definition of $m$ at the discontinuity points. 

\begin{figure}
\center{\includegraphics[width=8cm]{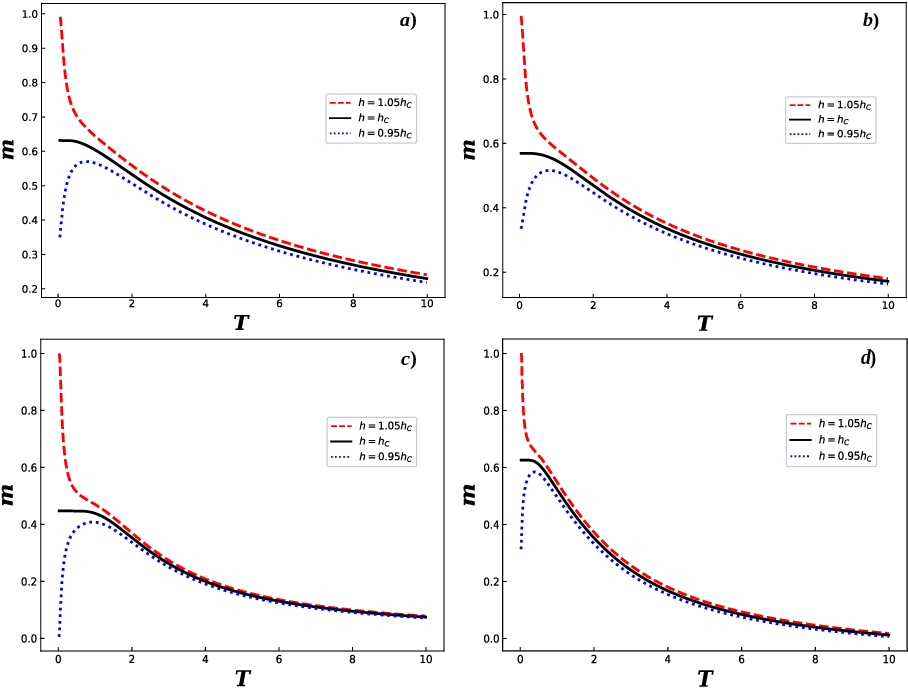}}
\caption{The dependencies of magnetization $m$ on temperature $T$ as the function of $h$ in the vicinity of $h_c$. 
The dotted curve on lower panel corresponds to smaller value of $h$ ($h=0.95h_c$), dashed curve corresponds to larger value of $h$ ($h=1.05h_c$). 
a) $\{J_d=-1$, $J=-1$, $J_t=-1\}$; b) $\{J_d=-1$, $J=-1$, $J_t=1\}$; c) $\{J_d=1$, $J=1$, $J_t=-1\}$;
d) $\{J_d=1$, $J=-1$, $J_t=-1\}$.
}
\label{fig6_8}
\end{figure}

Fig.~\ref{fig6_8} demonstrates the dependencies of magnetization $m$ on temperature $T$ as the function of $h$ in the vicinity of $h_c$. 
The dotted curve corresponds to smaller value of $h$ ($h=0.95 h_c$), dashed curve corresponds to larger value of $h$ ($h=1.05 h_c$) and
solid curve corresponds to $h=h_c$. 

\begin{figure}
\center{\includegraphics[width=8cm]{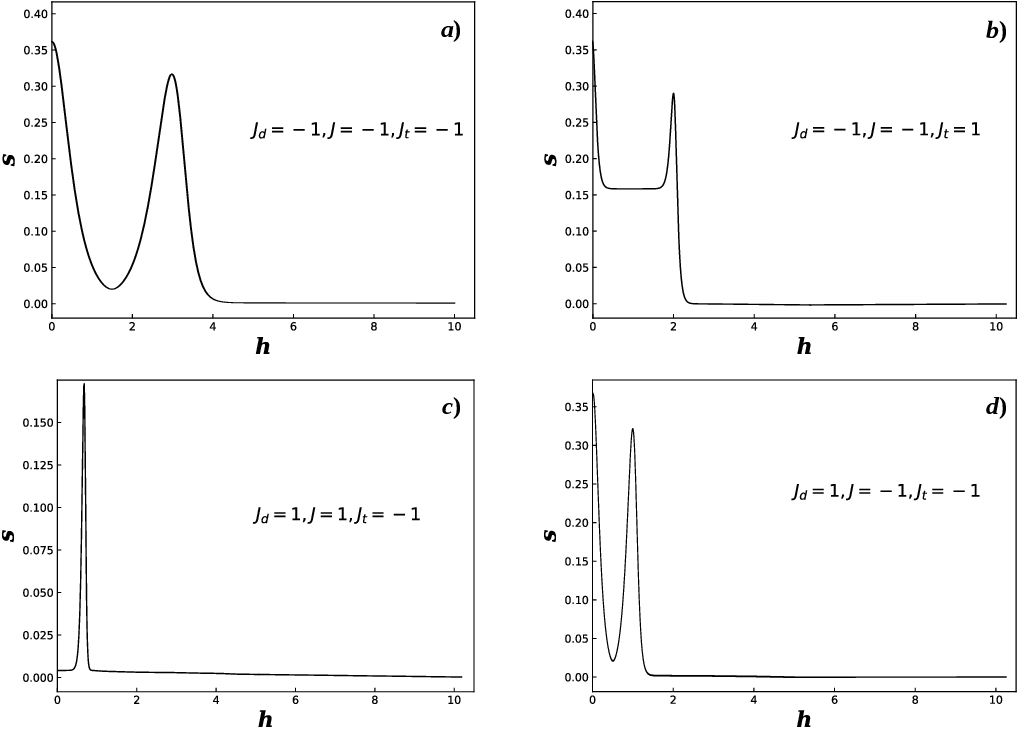}}
\caption{
Low temperature $T = 0.1$) field dependencies of the entropy $s$ at different values of exchange constants
a) $\{J_d=-1$, $J=-1$, $J_t=-1\}$; b) $\{J_d=-1$, $J=-1$, $J_t=1\}$; c) $\{J_d=1$, $J=1$, $J_t=-1\}$;
d) $\{J_d=1$, $J=-1$, $J_t=-1\}$.
}
\label{fig:s_h_low_T}
\end{figure}

\begin{figure}
\center{\includegraphics[width=8cm]{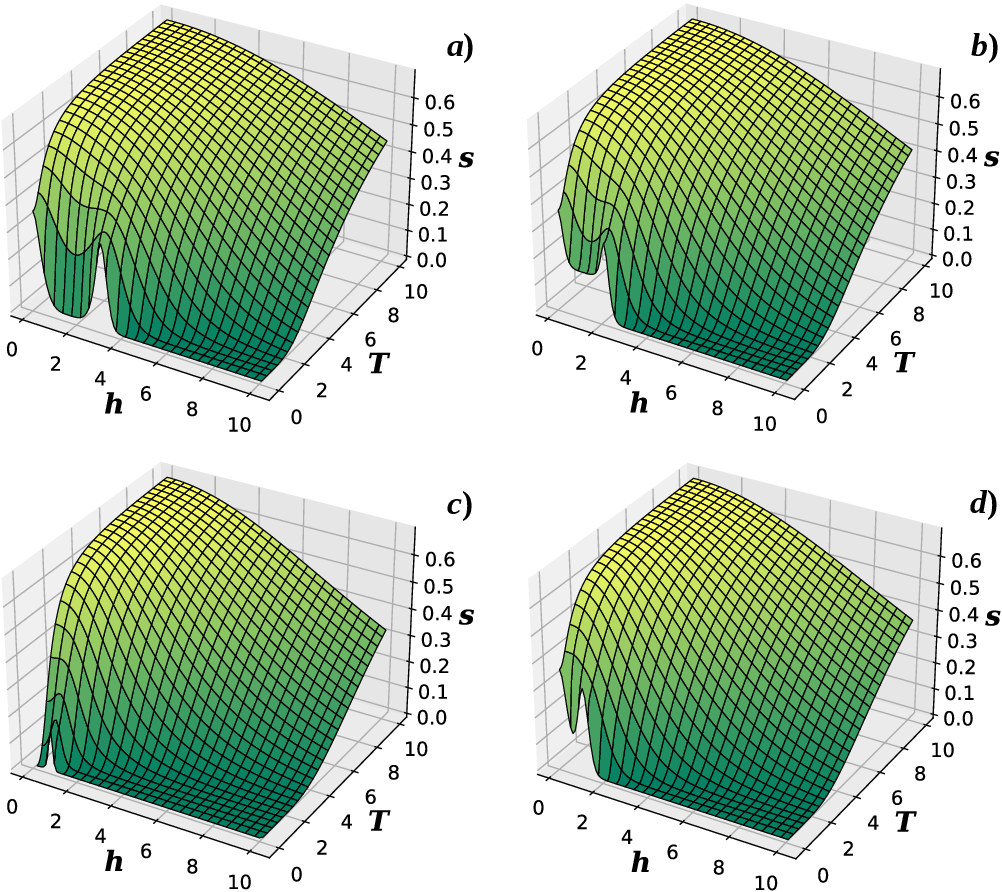}}
\caption{The dependence of entropy $s$ of field $f$ and temperature $T$ for different values of exchange constants. 
a) $\{J_d=-1$, $J=-1$, $J_t=-1\}$; b) $\{J_d=-1$, $J=-1$, $J_t=1\}$; c) $\{J_d=1$, $J=1$, $J_t=-1\}$;
d) $\{J_d=1$, $J=-1$, $J_t=-1\}$.}
\label{fig:s_h_and_T}
\end{figure}

The low-temperature ($T=0.1$) dependencies of entropy per particle $s$ as the function of $h$ are presented in Fig.~\ref{fig:s_h_low_T}.
The dependencies of $s$ as the function of $h$ and $T$ are shown in Fig.~\ref{fig:s_h_and_T}.
As it seen, the entropy demonstrates peaks at the quantum phase transitions fields $h=h_c$.
Besides, for the cases {\it a}, {\it b} and {\it d} one can see a peak at $h=0$, corresponding to an additional degeneration with respect to the states 
with $m=\pm m_0$ magnetization. The ground state values of $s$ corresponding to these peaks are presented in paragraph \ref{sec:gs}.

\section{The ground state properties of decorated Ising  chain.}
\label{sec:gs}
Using regular combinatorial considerations we will calculate the ground state entropy $s$ and magnetization $m$.
The ground state entropy (per particle) is:
\begin{equation}
s =\lim_{N_t\to\infty} \frac{1}{3N_t} \ln\left(\Omega_{N_t} \right)
\label{gs_entropy}
\end{equation}
\noindent Here $N_t$ is the number of triangles (the total number of particles is, hence, $N=3N_t$),  $\Omega_{N_t}$ is the number of configurations with minimal energy as the function of $N_t$. 
The corresponding  zero-temperature magnetization (per particle) is: 
\begin{equation}
m=\lim_{N_t\to \infty} \frac{1}{3N_t}\frac{M_{\Omega_{N_t}}}{\Omega_{N_t}}=\lim_{N_{t}\to \infty}\frac{m_{\Omega_{N_t}}}{\Omega_{N_t}}
\label{gs_magnetization}
\end{equation}
\noindent where $M_{\Omega_{N_t}}$ is the total magnetizations over all the ground state configurations as the function of 
$N_t$ and $m_{\Omega_{N_t}}=M_{\Omega_{N_t}}/(3N_t)$ is corresponding specific total magnetization.

\subsection{$\{J_d=-1$, $J=-1$, $J_t=-1\}$.} 

\begin{itemize}
\item $h=0$. For convenience we will consider $h=0$ as $\lim_{h\to +0}$. It allows us to remove degeneration related to the configurations of spins in triangles, where 2 spins have direction 
``$\uparrow$'' and 1 spin has direction ``$\downarrow$'' and vice verse, 1 spin ``$\uparrow$'' and 2 spins ``$\downarrow$'' (it is evident that the configurations triangles where all spins have the same direction have higher energy). 
In other words, it allows us to remove degeneration related to configurations of triangles with magnetization $m=1$ and  $m=-1$. Hence, all the ground state configurations must be constructed from triangles with $m=1$ and, thus, the ground state magnetization per spin 
$$m=1/3.$$
\noindent This value coincides with the magnetization $m$ obtained numerically (see Fig.~\ref{fig:m_h_low_T}a). 

The number of the ground state configurations as the function of $N_t$ is:
$$\Omega_{N_t}=4,10,28,82,244,730,\ldots, (1+3^{N_t}).$$

\noindent Hence, the corresponding ground state entropy per spin is:
$$s=\lim_{N\to\infty}\frac{1}{3N_t}\ln {\Omega_{N_t}} = \frac{1}{3}\ln(3)\approx 0.366.$$
\noindent As it seen, this value is close to obtained numerically (see Fig.~\ref{fig:s_h_low_T}a). 

\item $h = h_c=3$ (see Table \ref{Table1}).

$$\Omega_{N_t} = 3, 7, 18, 47, 123, 322, 843, \ldots$$

\noindent $\Omega_{N_t}$ in this case is the even Lucas numbers $\Omega_{N_t}=L_{2N_{t}}$:

\begin{equation}
L_{n}=\left(\frac{1+\sqrt{5}}{2}\right)^{n}+\left(\frac{1-\sqrt{5}}{2}\right)^{n},
\label{Lucas_Numbers}
\end{equation}

$$L_{n}=1,3,4,7,11,18,29,47,76,123,\ldots$$

The numerical sequence for $3m_{\Omega_{N_{t}}}$ represents the odd
Fibonacci numbers: $3m_{\Omega_{N_{t}}}=F_{2N_{t}+3}$.
\begin{equation}
F_{n}=\frac{1}{\sqrt{5}} \left[
\left(\frac{1+\sqrt{5}}{2}\right)^{n}-\left(\frac{1-\sqrt{5}}{2}\right)^{n}\right],
\label{Fibonacci_Numbers}
\end{equation}

$$F_{n}=1,1,2,3,5,8,13,21,34,55,\ldots.$$

\noindent Hence, 
$$s=\lim_{N_{t}\to \infty} \frac{1}{3N_{t}}\ln L_{2N_{t}}=$$
$$\lim_{N_{t}\to \infty}\frac{1}{3N_{t}} \ln \left[
\left(\frac{1+\sqrt{5}}{2}\right)^{2N_{t}}+\left(\frac{1-\sqrt{5}}{2}\right)^{2N_{t}}\right]=$$
$$\frac{2}{3}\ln \left(\frac{1+\sqrt{5}}{2}\right) \approx 0.321$$
\noindent (see Fig.~\ref{fig:s_h_low_T}a).

The corresponding magnetization is:

$$m=\frac{1}{3}\lim_{N_{t}\to \infty}\frac{F_{2N_{t}+3}}{L_{2N_{t}}}=$$
$$\frac{1}{3\sqrt{5}}\lim_{N_{t}\to \infty} \frac{\left(\frac{1+\sqrt{5}}{2}\right)^{2N_{t}+3}-\left(\frac{1-\sqrt{5}}{2}\right)^{2N_{t}+3}}
{\left(\frac{1+\sqrt{5}}{2}\right)^{2N_{t}}+\left(\frac{1-\sqrt{5}}{2}\right)^{2N_{t}}}=$$
$$\frac{1}{3}\left(1+\frac{2}{\sqrt{5}}\right) \approx 0.631$$
\noindent (see Fig.~\ref{fig:m_h_low_T}a).
\end{itemize}

\subsection{$\{J_{d}=J=-1,J_{t}=+1\}$.} 
\begin{itemize}

\item $h=0$. As far as $J_{d}=J=-1$ we have the same situation as in case a) in the ground state all the triangles must have magnetization $1$,
and, hence, the ground state magnetization per spin 
$$m=1/3$$ 

\noindent (see Fig.~\ref{fig:m_h_low_T}b).

The number of the ground state configurations is:

$$\Omega_{N_{t}}=\left(-1\right)^{N_{t}}+3^{N_{t}}$$

\noindent Hence 
$$s=\lim_{N_{t}\to \infty}\frac{1}{3N_{t}} \ln\left(\left(-1\right)^{N_{t}}+3^{N_{t}}  \right)=$$
$$\frac{1}{3}\ln\left(3\right) \approx 0.366$$
\noindent (see Fig.~\ref{fig:s_h_low_T}b).

\item $h=h_c=2$ (see Table \ref{Table1}).

$$\Omega_{N_{t}}=2,6,14,34,82,198,478,1154,\ldots$$
$$3m_{\Omega_{N_{t}}}=4,10,24,58,140,338,816,1970,\ldots$$

In this case $\Omega_{N_{t}}$ is the associated Pell numbers (or Pell-Lucas numbers)
$\Omega_{N_{t}}=Q_{N_{t}}$, determined by the formula:

$$Q_{n}=\left(1+\sqrt{2}\right)^{n}+\left(1-\sqrt{2}\right)^{n}$$

The sequence for $3m_{\Omega_{N_{t}}}$ represent the twice
Pell numbers: $3m_{\Omega_{N_{t}}}=2P_{N_{t}+1}$.

$$P_{n}=\frac{1}{2\sqrt{2}} \left[\left(1+\sqrt{2}\right)^{n}-\left(1-\sqrt{2}\right)^{n} \right]$$

The ground state entropy is:

$$s=\lim_{N_{t}\to \infty} \frac{1}{3N_{t}}\ln  Q_{N_{t}}=$$
$$s=\lim_{N_{t}\to \infty}\frac{1}{3N_{t}} \ln \left[
\left(1+\sqrt{2}\right)^{N_{t}}+\left(1-\sqrt{2}\right)^{N_{t}}
\right]$$
$$s=\frac{1}{3}\ln\left(1+\sqrt{2} \right) \approx 0.294$$
\noindent (see Fig.~\ref{fig:s_h_low_T}b).

The corresponding magnetization equals:

$$m=\frac{1}{3}\lim_{N_{t}\to \infty}\frac{2P_{N_{t}+1}}{Q_{N_{t}}}$$

$$m=\frac{1}{3\sqrt{2}}\lim_{N_{t}\to \infty}\frac{
\left(1+\sqrt{2}\right)^{N_{t}+1}-\left(1-\sqrt{2}\right)^{N_{t}+1}
} { \left(1+\sqrt{2}\right)^{N_{t}}+\left(1-\sqrt{2}\right)^{N_{t}}}$$

$$m=\frac{1+\sqrt{2}}{3\sqrt{2}} \approx 0.569$$
\noindent (see Fig.~\ref{fig:m_h_low_T}b).

\end{itemize}

\subsection{ \{$J_{d}=J=+1,J_{t}=-1$\}.} 

\begin{itemize}
\item $h=0$ In this case the ground state is the alternating sequence of triangles where all the spins in each triangle either oriented 
``$\uparrow$'' or ``$\downarrow$'.
In some case this is one-dimensional Ising chain with antiferromagnetic exchange $J_t$ and effective spins, formed by 3 parallel spin in each triangle.
Hence, $m=0$ and  $s=0$ (see  Fig.~\ref{fig:m_h_low_T}c and  Fig.~\ref{fig:s_h_low_T}c).

\item $h=h_c=2/3$ (see Table \ref{Table1}).

$$\Omega_{N_{t}}=1,3,4,7,11,18,29,47,76,123,\ldots$$
$$m_{\Omega_{N_{t}}}=1,1,2,3,5,8,13,21,34,55,\ldots$$

The number of the ground state configurations $\Omega_{N_{t}}$ represents the Lucas numbers (\ref{Lucas_Numbers}): $\Omega_{N_{t}}=L_{N_{t}}$. 
The numerical sequence for $m_{\Omega_{N_{t}}}$ represents the
Fibonacci numbers (\ref{Fibonacci_Numbers}): $m_{\Omega_{N_{t}}}=F_{N_{t}}$. Zero-temperature entropy per particle is, therefore:

$$s=\lim_{N_{t}\to \infty}\frac{1}{3N_{t}} \ln L_{N_{t}}=$$
$$\lim_{N_{t}\to \infty}\frac{1}{3N_{t}} \ln \left[
\left(\frac{1+\sqrt{5}}{2}\right)^{N_{t}}+\left(\frac{1-\sqrt{5}}{2}\right)^{N_{t}}
\right]=$$
$$s=\frac{1}{3}\ln \left(\frac{1+\sqrt{5}}{2}\right) \approx 0.160$$
\noindent (see Fig.~\ref{fig:s_h_low_T}c). Zero-temperature magnetization equals:

$$m=\lim_{N_{t}\to \infty}\frac{F_{N_{t}}}{L_{N_{t}}}=$$
$$\frac{1}{\sqrt{5}}\lim_{N_{t}\to \infty} \frac {
\left(\frac{1+\sqrt{5}}{2}\right)^{N_{t}}-\left(\frac{1-\sqrt{5}}{2}\right)^{N_{t}}
}{
\left(\frac{1+\sqrt{5}}{2}\right)^{N_{t}}+\left(\frac{1-\sqrt{5}}{2}\right)^{N_{t}}
}=$$
$$\frac{1}{\sqrt{5}} \approx 0.447$$

\end{itemize}

\subsection{\{$J_{d}=+1,J=-1,J_{t}=-1$\}.} 
\begin{itemize}
\item $h=0$.  The speculation, similar to cases a) and b) show that in the ground state all the triangles must have 2 spins oriented ``$\uparrow$'' and 1 spin oriented ``$\downarrow$''.
It means, that

$$m=1/3$$

\noindent (see  Fig.~\ref{fig:m_h_low_T}d).

$$\Omega_{N_{t}}=4,10,28,82,244,730,2188,6562,\ldots$$

The number of the ground state configurations is:

$$\Omega_{N_{t}}=1+3^{N_{t}}$$

Zero-temperature entropy per particle equals:

$$s=\frac{1}{3N_{t}}\ln\left( 1+3^{N_{t}}
\right) = \frac{1}{3}\ln\left(3\right) \approx 0.366$$

\noindent (see Fig.~\ref{fig:s_h_low_T}d).

\item $h=h_c=1$ (see Table \ref{Table1}).

$$\Omega_{N_{t}}=3,7,18,47,123,322,843,2207,\ldots$$
$$3m_{\Omega_{N_{t}}}=5,13,34,89,233,610,1597,\ldots$$

The number of the ground state configurations 
is the even Lucas numbers (\ref{Lucas_Numbers}): $\Omega_{N_{t}}=L_{2N_{t}}$.
The numerical sequence for $3m_{\Omega_{N_{t}}}$ represents the odd
Fibonacci numbers (\ref{Fibonacci_Numbers}): $3m_{\Omega_{N_{t}}}=F_{2N_{t}+3}$.


Zero-temperature entropy per particle equals:

$$s=\lim_{N_{t}\to \infty}\frac{1}{3N_{t}} \ln L_{2N_{t}}=$$
$$\lim_{N_{t}\to \infty}\frac{1}{3N_{t}} \ln \left[
\left(\frac{1+\sqrt{5}}{2}\right)^{2N_{t}}+\left(\frac{1-\sqrt{5}}{2}\right)^{2N_{t}}
\right]=$$
$$\frac{2}{3}\ln \left(\frac{1+\sqrt{5}}{2}\right) \approx 0.321$$

\noindent (see Fig.~\ref{fig:s_h_low_T}d). The corresponding magnetization is:

$$m=\frac{1}{3}\lim_{N_{t}\to \infty}\frac{F_{2N_{t}+3}}{L_{2N_{t}}}=$$

$$\frac{1}{3\sqrt{5}}\lim_{N_{t}\to \infty} \frac {
\left(\frac{1+\sqrt{5}}{2}\right)^{2N_{t}+3}-\left(\frac{1-\sqrt{5}}{2}\right)^{2N_{t}+3}
} {\left(\frac{1+\sqrt{5}}{2}\right)^{2N_{t}}+\left(\frac{1-\sqrt{5}}{2}\right)^{2N_{t}}}=$$
$$\frac{1}{3}\left(1+\frac{2}{\sqrt{5}}\right) \approx 0.631$$

\noindent (see  Fig.~\ref{fig:m_h_low_T}d).

\end{itemize}

\section{Results and Discussion}
We have studied thermodynamic properties of 1D frustrated Ising chain. Using Kramers-Wannier transfer matrix approach the exact expressions for free energy, magnetization and entropy were obtained. For several combinations of exchange constants the temperature and external magnetic field dependencies of the magnetization and entropy were calculated. For these combinations of exchange constants the quantum phase transitions were observed. The parameters of these transitions (i.e. critical fields) were established. Besides, we obtained analytic expressions for the ground state values of magnetization and entropy for the considered combinations of exchange constants. It was shown that the dependencies of the number of states with minimum energy (the degeneration of the ground state) as the function of the number of particles are expressed in terms of well-known numerical sequences - Lucas numbers and Pell numbers, which, in the limit of a large number of particles, are proportional to the powers of the golden and silver sections. Such dependencies of the ground state degeneration lead to nonzero (residual) values of entropy per particle.


\end{document}